\documentclass{aastex}

\shorttitle{HST Imaging of ESO\,410-G005}
\shortauthors{Karachentsev et al.}

\begin{document}

\title{Hubble Space Telescope Photometry of the Dwarf Spheroidal 
Galaxy ESO\,410-G005\altaffilmark{1}}


\author{Igor D.\ Karachentsev and Margarita E.\ Sharina} 
\affil{Special Astrophysical Observatory, Russian Academy
          of Sciences, N.\ Arkhyz, KChR, 357147, Russia}
\email{ikar@luna.sao.ru, sme@luna.sao.ru}
\author{Eva K.\ Grebel\altaffilmark{2}}
\affil{Department of Astronomy, University of Washington, Box 351580,
     Seattle, WA 98195, USA}
\affil{Max Planck Institute for Astronomy, K\"onigstuhl 17, D-69117 Heidelberg,
     Germany}
\email{grebel@astro.washington.edu}
\author{Andrew\ E.\ Dolphin}
\email{dolphin@noao.edu}
\affil{Kitt Peak National Observatory, National Optical Astronomy Observatories,
     P.O.\ Box 26732, Tucson, AZ 85726, USA}
\author{Doug\ Geisler} 
\email{doug@kukita.cfm.udec.cl}
\affil{Departamento de F\'{\i}sica, Grupo de Astronom\'{\i}a, Universidad 
     de Concepci\'on, Casilla 160-C, Concepci\'on, Chile}
\author{Puragra\ Guhathakurta\altaffilmark{3}}
\email{raja@ucolick.org}
\affil{UCO/Lick Observatory, University of California at Santa Cruz, 
     Santa Cruz, CA 95064, USA}
\author{Paul\ W.\ Hodge} 
\email{hodge@astro.washington.edu}
\affil{Department of Astronomy, University of Washington, Box 351580,
     Seattle, WA 98195, USA}
\author{Valentina E.\ Karachentseva} 
\email{vkarach@aoku.freenet.kiev.ua}
\affil{Astronomical Observatory of Kiev University, Observatorna 3, 254053, 
     Kiev, Ukraine}
\author{Ata\ Sarajedini}
\email{ata@astro.wesleyan.edu}
\affil{Astronomy Department, Wesleyan University, Middletown, CT 06459, USA}
\and 
\author{Patrick\ Seitzer}
\affil{Department of Astronomy, University of Michigan, 830 Dennison Building, 
     Ann Arbor, MI 48109, USA}
\email{seitzer@astro.lsa.umich.edu}


\altaffiltext{1}{Based on observations made with the NASA/ESA Hubble Space
Telescope.  The Space Telescope Science Institute is operated by the
Association of Universities for Research in Astronomy, Inc. under NASA
contract NAS 5-26555.  }
\altaffiltext{2}{Hubble Fellow}
\altaffiltext{3}{Alfred P.\ Sloan Research Fellow}

\begin{abstract}
We present {\it Hubble Space Telescope} WFPC2 imaging of the nearby 
low-surface-brightness dwarf spheroidal galaxy ESO~410-G005, which has been 
resolved into stars for the first time.
The resulting color-magnitude diagram for about 2500 stars shows a red giant
branch with a tip at $I = 22\fm4\pm0\fm15$, which yields
a distance of $D_{\rm MW}= 1.9\pm0.2$ Mpc.  ESO~410-G005 is found to be
metal-poor with a mean metallicity of $\langle$[Fe/H]$\rangle = (-1.8\pm0.4)$
dex estimated from its red giant branch.  Upper asymptotic giant branch stars
appear to be present near the center of the galaxy, indicative of a
substantial, centrally concentrated intermediate-age population, unless 
these objects are artifacts of crowding.  Previous studies did not detect 
ESO~410-G005 in H$\alpha$ or in H\,{\sc i}.  Based on our distance estimate, 
ESO~410-G005 is a probable member of the Sculptor group of galaxies.
Its linear separation from the nearest spiral,
NGC~55, is 230 kpc on the sky.  The deprojected separation ranges from 340
to 615 kpc %is $(340\pm250)$ or $(615\pm160)$ kpc 
depending on the assumed distance of NGC\,55.  The deprojected distance from
the Sculptor group spiral NGC\,300 is ($385\pm200$) kpc.  ESO~410-G005
appears to be a relatively isolated dSph within the Sculptor group. 
Its absolute magnitude,
$M_{V,0} = (-12\fm1\pm0\fm2)$ mag, its central surface brightness,
$\mu_{V,0} = (22.7\pm0.1)$ mag arcsec$^{-2}$,
and its mean metallicity, $[Fe/H] = (-1.8\pm0.4)$ dex follow the trend observed
for dwarf galaxies in the Local Group.
\end{abstract}


%% KEYWORDS TO BE CHECKED!


\section{Introduction \label{Sect_Intro}}

The most numerous morphological type of galaxy among the 34 members
of the Local Group are the dwarf spheroidal (dSph) galaxies.  At present
there are 18 dSph galaxies known within D = 1 Mpc (van den Bergh 1999,
Karachentsev \& Karachentseva 1999,
Whiting et al.\ 1999).  Four of these were only discovered within the 
last two years.  Some dSphs are seen in the direction of nearby groups
(M81, Centaurus~A), but only for two dSph systems, BK5N and F8D1, 
membership in the M81 group was confirmed by direct distance
measurements based on their resolved stellar populations (Caldwell et al.\
1998). 

The Local Group dSph galaxies show well-established relationships between
their global parameters such as absolute magnitude, central surface 
brightness and metal abundance.  However, it is still unknown whether these 
relationships are the same for dwarf galaxies in different environments.  
In this paper we describe the structural parameters and the properties of 
the stellar populations of a dSph galaxy in the region of the nearby,
loose Sculptor group that is not
in the immediate vicinity of a more massive galaxy. 
This dSph galaxy is ESO~410-G005 (Lauberts 1982), also known as 
FG~11 (Feitzinger \& Galinski 1985), AM~0013-322 (Arp \& Madore 1987),
PGC~1038 (Paturel et al.\ 1989), and kk~003 (Karachentseva \& Karachentsev 
1998).  ESO~410-G005 was classified as dwarf elliptical (dE) by
Feitzinger \& Galinski (1985), who nevertheless noted its knotty structure.
Karachentseva \& Karachentsev (1998) considered
ESO~410-G005 to be a Sph/Irr due to the presence of knots on its eastern side.
Miller (1996) imaged it in the H$\alpha$ line and did not find any emission.
Radio surveys by 
Longmore et al.\ (1982), Maia et al.\ (1993), C\^ot\'e et al.\ (1997), and
Huchtmeier et al.\ (2000) did not detect ESO~410-G005 in the H\,{\sc i} 21\,cm
line.  Huchtmeier et al.\ (2000) found
an upper flux limit $S < 24$ mJy, which indicates a low amount of gas as
typical for dSph galaxies.  
A Digital Sky Survey image of ESO~410-G005 is shown in Fig.\ 1. The galaxy
has a slightly granulated appearance in the image with a total dimension of
$1\farcm5\times1\farcm3$.

\section{{\em HST} WFPC2 photometry}

In order to study the stellar populations of ESO~410-G005 and to measure its
distance, we imaged this dwarf galaxy with the Wide Field and Planetary Camera 2
(WFPC2) aboard the {\em Hubble Space Telescope} (HST).  The data were
obtained on 1999 August 23 with exposure times of 600~s in the F606W 
and F814W filters.
These data were taken as part of an {\em HST} 
snapshot survey (program GO~8192, PI: Seitzer) of nearby dwarf galaxy
candidates from the list of Karachentseva \& Karachentsev (1998).
Fig.\ 2 shows an image of ESO~410-G005
(both filters combined). The galaxy was centered on the WF3 chip. After
removing cosmic ray hits we measured instrumental magnitudes using the
MIDAS implementation of the DAOPHOT crowded-field photometry program
(Stetson et al. 1990). The standard DAOPHOT II/ALLSTAR procedure was
used for automatic star finding and then measure stellar magnitudes by
fitting a point- spread function (PSF) in each filter for each WF chip.
The PSF photometry was made for about 6200 stars in both filters with
an aperture radius of 1.5 pixels. The F606W and F814W instrumental
magnitudes were first transformed to the Holtzman et al. (1995) 0$\farcs$5
aperture magnitudes by determining the aperture correction that need
to be applied to the PSF magnitudes. Typically 10 -- 50 of the brightest,
most isolated stars spread across each WF image were used. We found the
mean aperture corrections to be in the range of 0.43 -- 0.46 mag (F606W)
and 0.49 -- 0.52 mag (F814W).

We then used equations 1a, 1b, and 3 from Whitmore et al.\ (1999) to correct
the magnitudes for the charge-transfer efficiency (CTE) loss, which
depends on the X- and Y- positions, the background counts, the brightness
of the stars and the time of the observations. For our data
with typical background counts of ~65 e (F606W) and 35 e (F814W)
the mean CTE correction makes a star brighter by 0.11 mag and bluer by
0.02 mag relative to the uncorrected magnitudes.

Finally, the F606W and F814W instrumental magnitudes were converted to
the standard $V, I$ system following the ``synthetic'' transformations of
Holtzman et al.\ (1995). We used the parameters of transformations from their
Table 10 taking into account different relations for blue and red stars
separately. Because we used the non-standard $V$ filter F606W instead of
F555W, the resulting $I$ and especially $V$ magnitudes may contain systematic
errors. However, when comparing our \{F606W, F814W\} photometry of other 
snapshot targets with ground-based $V, I$ photometry we find that 
the transformation uncertainties, $\sigma(I)$ and $\sigma(V-I)$,
are within 0.1 mag for stars with colors of  $0 < (V-I) < 2$.

Finally, objects with goodness of fit parameters
$\mid$SHARP$\mid$ $> 0.3$, $\mid$CHI$\mid >$ 2, and
$\sigma (V) > 0.2$ mag were excluded. The resulting color-magnitude diagrams
(CMDs) in $I$, $V-I$ for $\sim$2500 stars are presented in Fig.\ 3.

\section{Color-magnitude diagram \label{Sect_CMD}} 
                                   
   The first panel of Fig.\ 3 represents a CMD for
the central WF3 field, which covers the main body of ESO~410-G005. The next 
panel shows the CMD for the neighboring regions in the southern half of WF2 
and the eastern half of WF4 (the ``medium'' field), and the last panel 
contains stars found in the remaining outer halves 
of WF2 and WF4.  In the central field the number of stars increases abruptly
at $I = 22\fm4$, which we interpret as the tip of the red giant branch (TRGB).
The same feature is seen also in the medium field, which corresponds to the
outer regions of the galaxy.  
No bright blue stars with $V-I < 0\fm7$ are present in the medium field. 
This allows us to estimate a lower limit for the most recent star formation 
episode.  Using the Bertelli et al.\
(1994) isochrones and adopting the distance and mean abundance
from the next sections for ESO~410-G005, the absence of bright blue stars 
indicates that no star formation has occurred
in the galaxy halo for the last 300~Myr. 

A significant number of red stars with  $I < 22\fm4$ is evident in the central
field.  These are probably upper asymptotic giant branch (AGB) stars
indicating the presence of an intermediate-age population ($\la 10$ Gyr).
One can also see many faint, bluish stars with $(V-I)<0\fm5$
in the central field. These stars 
may be indicative of a young population in the core
of ESO~410-G005.  Centrally concentrated intermediate-age populations and 
in some cases young populations have been found in a number of other
dSphs as well (see Grebel 1999 for a review). 
However, the blue wing of the CMD and the candidate
AGB stars may instead be artifacts of  
crowding in the central part of the galaxy (e.g., Grillmair et al.\ 1996).

  To check the significance of stellar crowding we carried out simulations
in which we added artificial stars in the central ($450\times450$ pixels) 
part of the galaxy using the ADDSTAR routine. For each pair of $I$ magnitude 
and $V-I$ color listed in the left side
of Table 1 we inserted one hundred randomly distributed artificial
``stars'', and then
applied the same DAOPHOT detection and photometry algorithms as before.  The
results are presented in Fig.\ 4 and Table 1. We can conclude from these
simulations that at $I_{lim} = 24\fm8$ the detection rate has dropped to 50\%.
The scatter of colors for the detected stars increases towards faint
magnitudes, following roughly the RGB ridge line. A noticeable fraction of
artificial stars is situated above the TRGB, but a few of the simulated
stars have colors bluer than $0\fm5$. Therefore, the effect of stellar
blending in the central galaxy region may partially be responsible for
the presence of stars with $I \leq 22$ mag in the CMD of ESO~410-G005.
The results in Table 1 indicate also that the position of the TRGB
shifts to a brighter (~0.10 mag) and bluer (~0.04 mag) magnitude due
to stellar crowding. The same effect has been shown by Madore \& Freedman
(1995). Thus, in transforming the WF3 photometry to the standard $V, I$
system we applied a zero point shift of $\delta I = 0.10$ mag and
$\delta(V-I) = 0.04$ in the sense of fainter magnitudes and redder colors,
which reduced a slight difference between the TRGB positions
in the central and the medium parts of ESO 410-G005.

\section{Distance \label{Sect_Dist}}

According to Da Costa~\& Armandroff (1990), the TRGB can be assumed 
to be at $M_I = -4\fm05$ for metal-poor systems. We find the apparent magnitude 
of the TRGB of ESO~410-G005
to be $I_{\rm TRGB} = 22\fm4\pm0\fm15$. The one-sigma error
here is determined roughly by estimating a scatter in position of the sharp
rise in the $I$-band luminosity function of the galaxy (Fig.\ 5) under
different manner of binning. With a Galactic extinction along the line
of sight toward ESO~410-G005 of $A_I = 0\fm03$ (Schlegel et al.\ 1998)
this yields a distance
modulus of $(m-M)_0= 26\fm42\pm0\fm20$ or $D = (1.92\pm0.19)$ Mpc.
The quoted errors include uncertainties of the
synthetic transformation ($\sim$0.10) and crowding effects ($\sim$0.10). 
The solid lines in Fig.\ 3a are
globular cluster fiducials from Da Costa~\& Armandroff (1990), which were 
reddened and shifted to the galaxy's distance.
The fiducials cover a range of [Fe/H] values
(from left to right): $-2.2$ dex (M15), $-1.6$ dex (M2), and $-1.2$ dex 
(NGC\,1851).

\section{Metal abundance}

With knowledge of the distance modulus of ESO~410-G005 we can estimate its 
mean metallicity 
from the mean color of the red giant branch (RGB)
measured at an absolute magnitude
$M_I= -3\fm5$, as described by Da Costa~\& Armandroff (1990). Based on a
Gaussian fit to the color distribution of the giant stars in the range
$22\fm7 < I < 23\fm1$ we derive a mean dereddened color of the RGB stars of
$(V-I)_{0,-3.5} = 1\fm30\pm0\fm03$. The reddening towards ESO~410-G005 is
$E(V-I) = 0.018$. Following Lee et al.\ (1993)
this yields a mean metallicity  $\langle$[Fe/H]$\rangle$ = 
($-1.84\pm0.12$) dex. However, this uncertainty must be considered an intrinsic
error. Further error sources come from the effect of stellar crowding
($\sim$0.04) and an uncertainty of the transformation zeropoint for the $V$
magnitudes, and hence the $(V-I)$ colors ($\sim$0.10). Added in quadrature 
they give a total uncertainty $\sigma(V-I) = 0.11$ or $\sigma$[Fe/H] = 0.4 dex.

\section{Integrated properties}

The radial distribution of the surface brightness in the $V$ band 
averaged over azimuth is shown in the right panel of Fig.\ 6. The left panel
reproduces the radial variation of the $(V-I)$ color also averaged in azimuth.
In a distance interval $10\arcsec< R < 35\arcsec$ the surface brightness
profile is well approximated
by an exponential fit with a scale length $h = 14\farcs5\pm0\farcs5$. The
observed central surface brightness is 
$\mu_{V,0} = (22.7\pm0.1)$~mag~arcsec$^{-2}$.
The mean galaxy color becomes slightly redder
towards the periphery of the galaxy. 
This may be caused by a radial age gradient in ESO~410-G005 (see Section
\ref{Sect_CMD}).
The galaxy's total color index, $(V-I)_T = 0\fm90\pm0\fm05$, is determined as
the difference of integral magnitudes in each band within a radius
of 35$\arcsec$.  This value is in excellent agreement with the result 
by Miller (1994) of $(V-I)_T = 0\fm89\pm0\fm07$ from ground-based CCD 
photometry.  Furthermore, Miller (1994) found $B_T = 15\fm12$ and
$(B-V)_T = 0\fm77\pm0\fm05$, and Lauberts~\& Valentijn (1989) measured 
$(B-R) = 0\fm81\pm0\fm09$ inside the standard
25 mag arcsec$^{-2}$ isophote.  With the distance and extinction estimates from
Section \ref{Sect_Dist}, we can derive the integrated absolute magnitude
of ESO~410-G005:  $M_{B,0} = -11\fm36\pm0\fm2$ and $M_{V,0} =
-12\fm07\pm0\fm2$.  

A summary of the basic parameters of ESO~410-G005 is given in Table 2. The data
in the first six lines are from NASA's Extragalactic Database (NED).  The
total $B$ and $V$ magnitudes were adopted from Miller (1994), while 
the other listed parameters, except for the extinction value, are from this 
paper.  The symmetric shape of the galaxy, its smooth surface brightness
profile, the reddish total colors, the
lack of an appreciable amount of neutral hydrogen, and the absence of a
significant young population favor the classification  
of ESO~410-G005 as a dwarf spheroidal galaxy.
 
The integrated absolute magnitude of ESO~410-G005 and its
standard linear diameter, 0.72 kpc, correspond to the parameters typical
for spheroidal companions of the Milky Way and M31. As seen from
Fig.\ 7, the derived parameters of
ESO~410-G005 follow the general relationships between central surface
brightness, $\mu_{V,0}$, absolute magnitude,
and the mean metallicity defined by Local Group
dwarfs (Caldwell et al.\ 1998, Grebel~\& Guhathakurta 1999).

We searched for globular clusters in ESO~410-G005 but found no candidates 
within the appropriate range of colors and magnitudes defined by
Milky Way globulars (Harris 1996).  This null result is not surprising
given the expected value of the specific frequency of globular clusters in 
low-luminosity dSph and dE
galaxies (Harris~\& van den Bergh 1981; Miller et al.\ 1998). 

On the eastern
side of ESO~410-G005 there is a group of diffuse, extended objects (Fig.\ 2).
These objects are the ``knots'' that had been detected earlier from the ground
(Section \ref{Sect_Intro}). Integral 
magnitudes for them as well as their colors, central surface brightnesses and 
half-light radii are given in Table 3. Two of the brightest objects have faint
external features resembling tidal tails (Fig.\ 8). Judging from their 
morphology and color these objects seem to be members of 
a group of background galaxies.

\section{Environmental status} 

ESO~410-G005 is located in the direction of the loose group of galaxies 
in Sculptor, very close to the SuperGalactic equator ($b_{SG} = -0.26\degr$). 
Several multiple galaxy systems situated at different distances overlap with 
each other along the line of sight in this complicated region (Jerjen et al.\ 
1998). In the wide vicinity of ESO~410-G005 there are 12 galaxies with 
distance estimates $D < 4$ Mpc or with radial velocities $V_0 < 300$ km 
s$^{-1}$, where $V_0$ is measured relative to the
barycenter of the Local Group (Karachentsev~\& Makarov 1996).  These galaxies
are listed in Table 4, which is an updated version of Table 5 from Jerjen et
al.\ (1998).  The sky distribution of these galaxies is shown in Fig.\ 9.
Large and small filled squares correspond to luminous and dwarf late-type
galaxies, respectively, and open squares indicate four dSph galaxies.
The two bright spiral galaxies NGC\,55 and NGC\,300 are both at the near  
side of the elongated Sculptor group.  It has been
suggested that NGC\,55 and NGC\,300 form a bound pair (Graham 1982, Pritchet et 
al.\ 1987, Whiting 1999).  The distance of NGC\,55 is poorly known.
Distance estimates for NGC\,55 range from
1.34 Mpc based on Carbon stars (Pritchet et al.\ 1987) to 1.66 Mpc from the
Tully-Fisher relation (Puche~\& Carignan 1988).  Davidge (1998) finds
NGC\,55 and NGC\,300 to be at a $7\fm5$ larger distance modulus than the LMC,
i.e., at $\sim 1.6$ Mpc.  The distance to NGC\,300 was measured via 
Cepheids (Freedman et al.\ 1992: ($2.1\pm0.1$) Mpc) and via the planetary
nebulae luminosity function (Soffner et al.\ 1996: ($2.4\pm0.4$) Mpc).
In projection ESO~410-G005 is 
closest to NGC\,55 (see Table 3 and Fig.\ 8; NGC\,7793 is much more distant).
The dwarfs 
ESO~410-G005 as well as PGC~1641 and possibly PGC~621 may be remote 
companions of the bright galaxy pair.  

ESO~410-G005 has a linear projected separation of ($230\pm10$) kpc from 
NGC\,55 (assuming that NGC\,55 is at the same distance as ESO~410-G005).
The deprojected distance of ESO~410-G005 ranges from 340 kpc to 
615 kpc, depending on whether Puche~\& Carignan's (1988) or 
Pritchet et al.'s (1987) distance to NGC\,55 is considered.  The formal errors
of these values are $\pm 250$ and $\pm 160$ kpc when taking the uncertainties
in the Galactocentric distance determinations to NGC\,55 and ESO~410-G005 
into account.  However, ESO~410-G005 cannot be closer to NGC\,55 than its 
linear projected separation, which gives a lower limit to their true
separation within the Sculptor group. 

The linear projected separation between ESO~410-G005 and NGC\,300 is
$\sim 330$ kpc if both galaxies were at the distance of ESO~410-G005.
The deprojected distance between these two objects is ($385\pm200$) kpc
when we adopt the Cepheid distance for NGC\,300.  As before the linear
separation gives a better lower limit to their true separation.

NGC\,55 has a 5--6 times lower 
luminosity and mass than M31 or the Milky Way, and NGC\,300 is 
slightly more luminous than NGC\,55.  All dSph 
companions of the two large Local Group spirals 
are concentrated within a radius of 300 kpc. 
There is only one case of an isolated dSph in the Local
Group known, namely Tucana.  (The recently discovered dSph Cetus is 
close to both WLM and IC\,1613).
Considering the lower mass of NGC\,55 relative to M31 or the Milky Way, 
ESO~410-G005 seems to be located between NGC\,300 and NGC\,55 and may be 
a relatively isolated dSph within the Sculptor group.  A more accurate
distance determination for NGC\,55 is required to determine its location
with respect to ESO~410-G005.

\section{Concluding remarks}

Based on $V,I$ images obtained with the {\it Hubble Space Telescope} we have
resolved the dwarf spheroidal galaxy ESO~410-G005 into stars.  Knotty 
structures
in this galaxy visible in ground-based images are identified as a probable
distant background cluster of galaxies.  The CMD for $\sim$2500
stars shows the presence of a red giant branch with $I_{\rm TRGB} = 
22\fm4\pm0\fm15$, which yields a true distance modulus of $26\fm42\pm0\fm2$
or $D = 1.92\pm0.19$ Mpc.  The galaxy's global parameters, namely the
absolute magnitude of $M_B = - 11\fm3$, the standard linear diameter of 
$D_{25}$ = 0.72 kpc, the central surface brightness of 
$\mu_{V,0} = 22.7$ mag arcsec$^{-2}$, and the  mean
metal abundance $\langle$[Fe/H]$\rangle$ = $-1.8$ dex
are within the range of properties observed for dSph companions of the 
Milky Way and M31.
 
In addition to RGB stars, the CMD for the central part of ESO~410-G005
shows a considerable number of brighter
red stars (candidate upper AGB stars) and also some blue stars.
There are two possible explanations of what these objects are: blends of fainter
stars due to crowding, or members of a centrally concentrated
intermediate-age population.
The radial variation of the integrated color
of ESO~410-G005 seems to be an independent line of evidence in
support a younger population in the center. 

ESO~410-G005 appears to lie between the closest Sculptor group spirals
NGC\,55 and NGC\,300.  The projected distance between ESO~410-G005 and
the moderately massive spiral NGC\,55 is 230 kpc (assuming that NGC\,55
were at the same distance as ESO~410-G005), and 330 kpc if NGC\,300 were
at the same distance as ESO~410-G005.
The deprojected separation between ESO~410-G005 and NGC\,55
ranges from 340 to 615 kpc depending on
the adopted distance of NGC\,55.  When NGC\,300's Cepheid distance
is adopted then its deprojected distance from ESO~410-G005 is ($385\pm200$) kpc.
All these deprojected distances are larger than the largest separations between
the dSph companions of M31 and the Milky Way in the Local Group, and NGC\,55
and NGC\,300 are less massive than the dominant Local Group spirals.
ESO~410-G005 may therefore be a 
relatively isolated dSph within the Sculptor group. 

Finally, we note that measuring the radial velocity of ESO~410-G005
in addition to the TRGB distance given here would allow one to constrain
its dynamical status within the Sculptor group.


%% Included in this acknowledgments section are examples of the
%% AASTeX hypertext markup commands. Use \url without the optional [HREF]
%% argument when you want to print the url directly in the text. Otherwise,
%% use either \url or \anchor, with the HREF as the first argument and the
%% text to be printed in the second.

\acknowledgements
Support for this work was provided by NASA through grant GO-08192.97A from
the Space Telescope Science Institute, which is operated by the Association
of Universities for Research in Astronomy, Inc., under NASA contract
NAS5-26555.  IDK, VEK, and EKG acknowledge partial support through the
Henri Chr\'etien International Research Grant administered by the American
Astronomical Society.  EKG acknowledges support by NASA through grant 
HF-01108.01-98A from the Space Telescope Science Institute.

This research has made use of the NASA/IPAC Extragalactic 
Database (NED) which is operated by the Jet Propulsion Laboratory, California 
Institute of Technology, under contract with NASA. We also used 
NASA's Astrophysics Data System Abstract Service and  the SIMBAD
database, operated at CDS, Strasbourg.  The Digitized Sky
Surveys were produced at the Space Telescope Science Institute under U.S.\
Government grant NAG W-2166. The images are based on
photographic data obtained using the UK Schmidt Telescope.
The UK Schmidt Telescope was operated by the Royal Observatory Edinburgh,
with funding from the UK Science and Engineering Research Council (later
the UK Particle Physics and Astronomy Research Council), until 1988 June,
and thereafter by the Anglo-Australian Observatory.




{}

%% Generally speaking, only the figure captions, and not the figures
%% themselves, are included in electronic manuscript submissions.
%% Use \figcaption to format your figure captions. They should begin on a
%% new page.

\clearpage

%% No more than seven \figcaption commands are allowed per page,
%% so if you have more than seven captions, insert a \clearpage

%% There must be a \figcaption command for each legend. Key the text of the
%% legend and the optional \label in curly braces. If you wish, you may
%% include the name of the corresponding figure file in square brackets.
%% The label is for identification purposes only. It will not insert the
%% figures themselves into the document.
%% If you want to include your art in the paper, use \plotone.
%% Refer to the on-line documentation for details.

\figcaption[kara.fig1.eps]{A Digital Sky Survey image of ESO\,410-G005
made from IIIaJ plate.
The field size is $4'$ on a side. North is up, and East is to the left.  The 
dark spot at the northern edge is an emulsion defect. 
\label{Fig_DSS}}

\figcaption[kara.fig2.eps]{WFPC2 image of ESO\,410-G005 produced by 
combining the two 600\,s exposures through the F606W and F814W
filters.  ESO\,410-G005 is centered in the WF3 chip (WF3-FIX mode). 
The arrow points to the North, while the other line points towards the
East.  Each line is $30''$ long.  At the eastern side of 
ESO\,410-G005 (bottom of image) is a group of distant background
galaxies.
\label{Fig_WFPC2}} 

\figcaption[kara.fig3.eps]{Color-magnitude diagram from the WFPC2 data 
of ESO\,410-G005. 
The three panels show diagrams based on stars within the central (WF3) field, 
the ``medium'' field (the neighboring halves of the WF2 and WF4 chips), and 
the outer field (remaining halves of the WF2 and WF4 chips).  Each of these
three fields covers an equal area of $800 \times 800$ pixels. The solid lines 
in the left panel show the mean loci of the red giant branches of globular 
clusters with different metallicities, [Fe/H]: M15 ($-2.17$ dex), M2 
($-1.58$ dex), 
and NGC~1851 ($-1.29$ dex) from left to right, based on Da Costa \& 
Armandroff (1990).
\label{Fig_CMDs}}

\figcaption[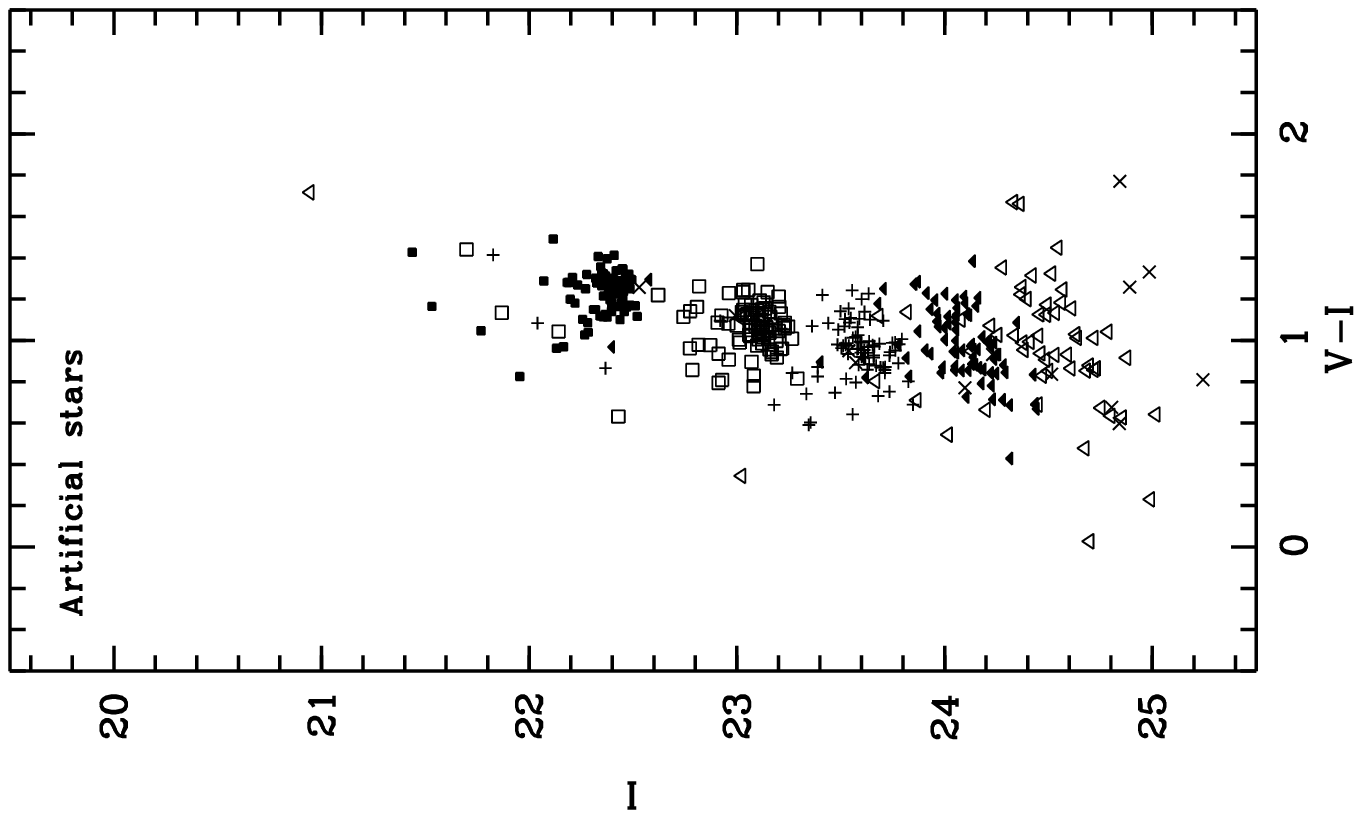]{Color-magnitude diagram for artificial stars.
The different symbols correspond to:
$I=22\fm41$ in Table 1 (filled squares), $I=23\fm12$ (open squares), 
$I=23\fm62$ (crosses ``+''), $I=24\fm12$ (filled triangles), $I=24\fm62$ 
(open triangles), and $I=25\fm12$ (crosses ``$\times$'').
\label{Fig_ART}}

\figcaption[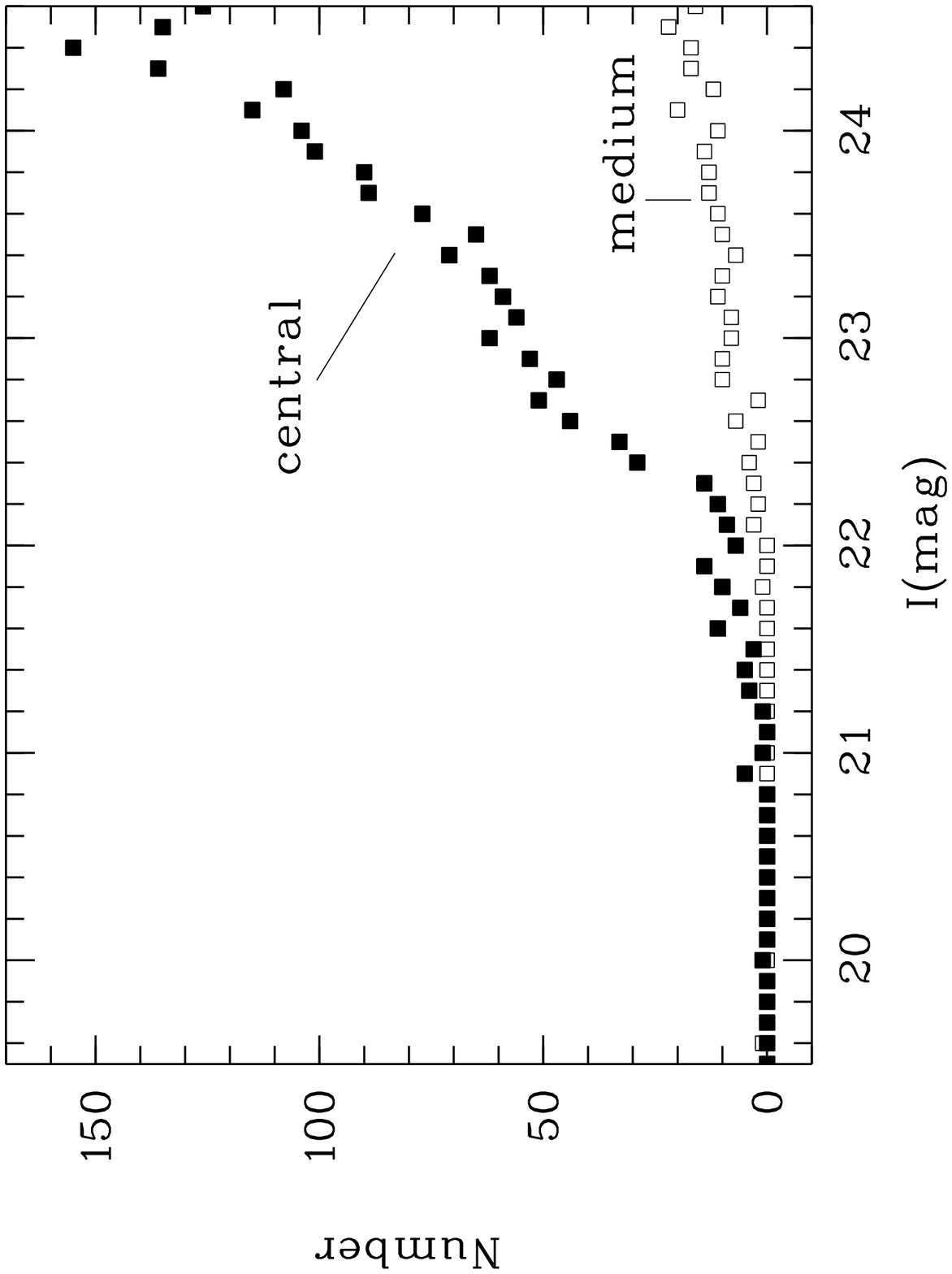]{$I$-band luminosity function of ESO\,410-G005.  The 
filled squares show the luminosity function of stars in the central
field (see Fig.\ 3), whereas open squares outline the luminosity
function of the neighboring fields.  The sharp rise in the luminosity 
function of the central field at $I = 22\fm4$ indicates the location of 
the tip of the red giant branch. 
\label{Fig_LUM}}

\figcaption[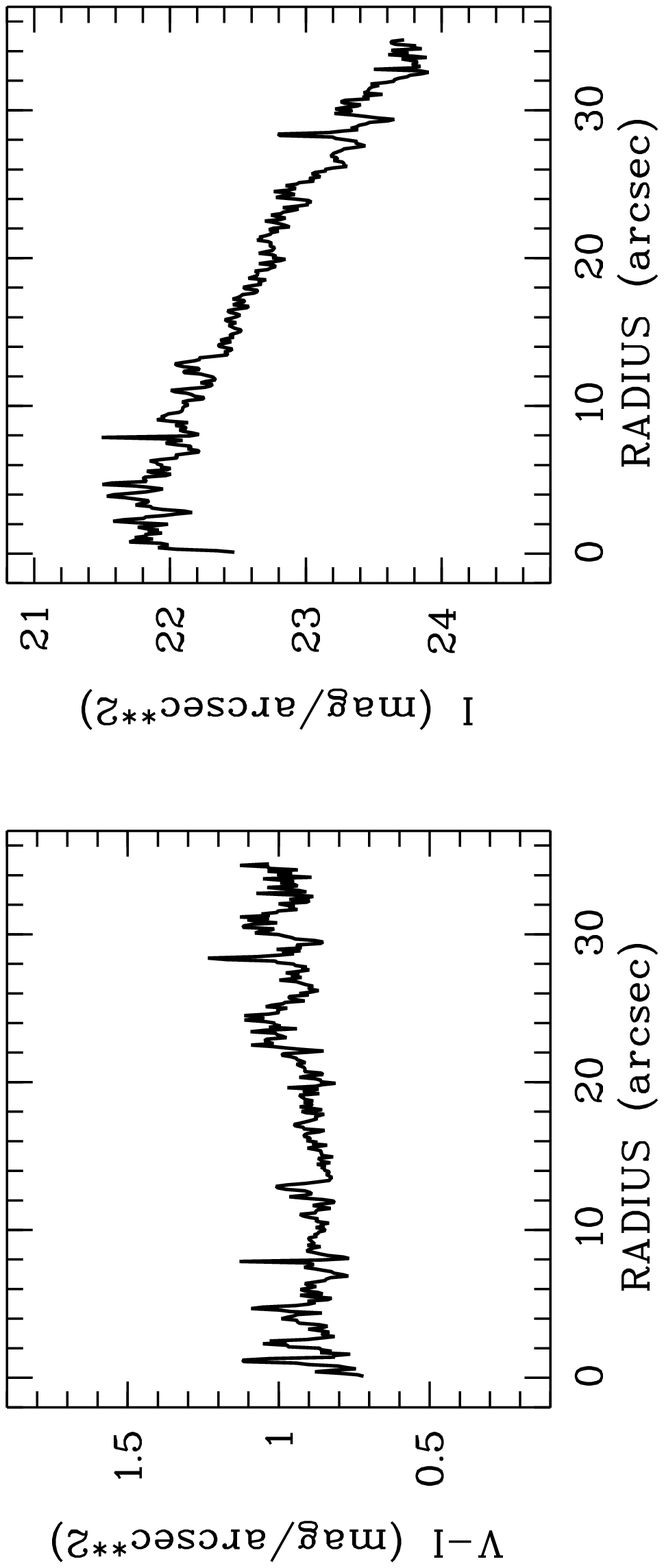]{Radial distribution of $V$-band surface brightness
(right) and $V-I$ color in ESO\,410-G005, averaged over azimuth
within circular annuli.
\label{Fig_surf}}

\figcaption[kara.fig7.ps]{
Central $V$ surface brightness, $\mu_{V,0}$,
versus mean metallicity, $\langle$[Fe/H]$\rangle$ (upper panel) and 
versus absolute $V$ magnitude, $M_V$ (lower panel)
for Local Group dwarf galaxies (filled
circles and diamonds; data from Grebel \& Guhathakurta 1999 and Caldwell
1999), M81 dwarfs (crosses; Caldwell et al.\ 1998), and for ESO\,410-G005
(open circle).
\label{Fig_FMS}}

\figcaption[kara.fig8.eps]{Enlargement of the eastern region of 
ESO\,410-G005 (WFPC2).  Several background galaxies are seen, some of which
seem to have extensions resembling tidal tails.  The global parameters
of these objects are given in Table 3.
\label{Fig_galaxies}}

\figcaption[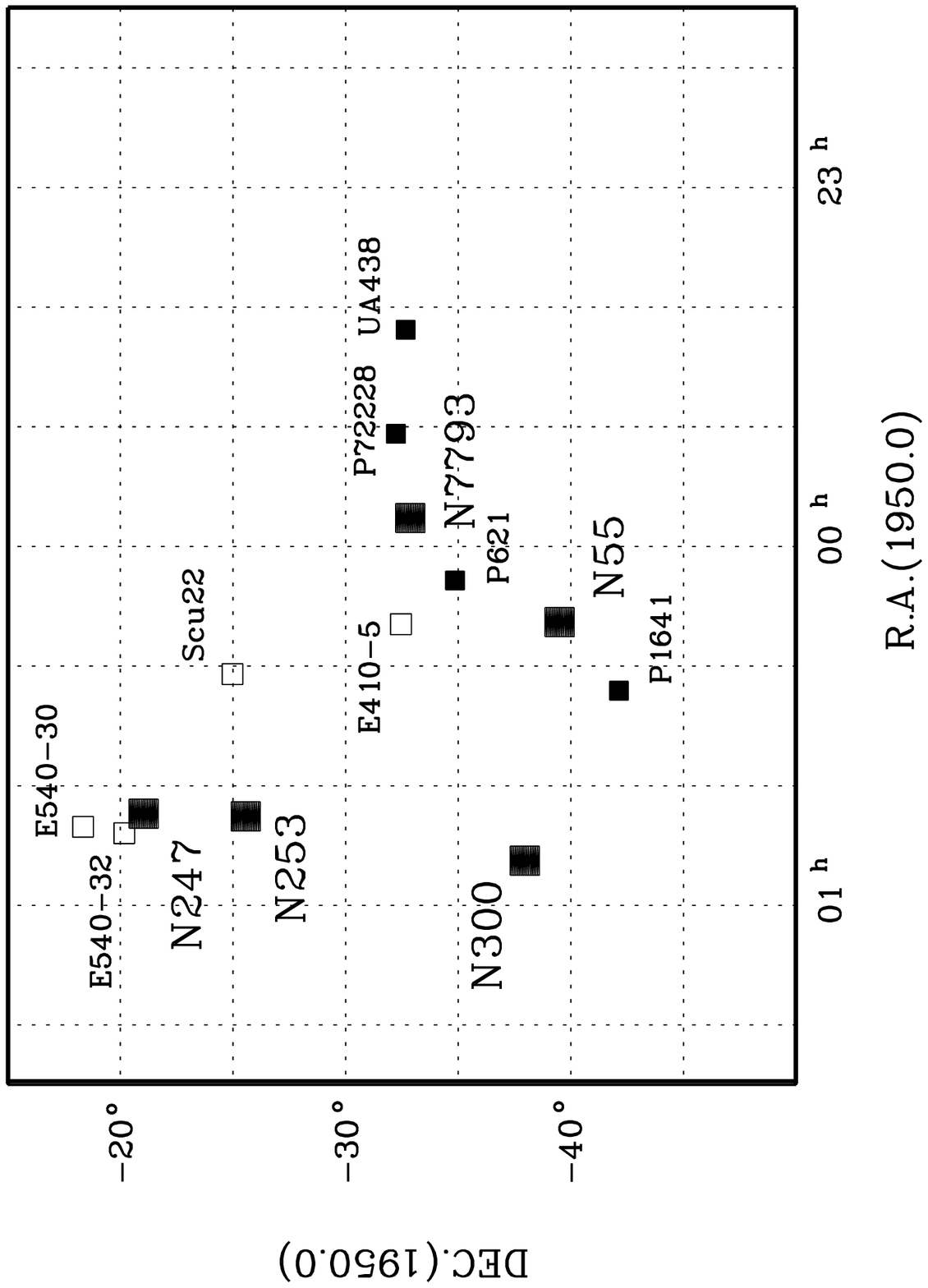]{Sky distribution of nearby galaxies in the
direction of the Sculptor group with known distances $D< 4$ Mpc or corrected 
radial velocities $V_0 < 300$ km s$^{-1}$. Small squares indicate dwarf
spheroidal (open symbols) and irregular (filled symbols) galaxies. Large 
squares
correspond to luminous galaxies with NGC numbers.  Note that the bright
spiral NGC\,7793, which is seen in close projection near ESO\,410-G005,
has a distance of $\sim 3.3$ Mpc placing it 1.4 Mpc behind ESO\,410-G005, 
while NGC\,55 is a true near neighbor.
\label{Fig_spatial}}

%% Tables should be submitted one per page, so put a \clearpage before

%% Two options are available to the author for producing tables:  the
%% deluxetable environment provided by the AASTeX package or the LaTeX
%% table environment.  Use of deluxetable is preferred.
%%

%% Three table samples follow, two marked up in the deluxetable environment,
%% one marked up as a LaTeX table.

%% In this first example, note that the \footnotesize command has been
%% used to shrink the table so it will fit on one page. Note also that
%% the \label command needs to be placed inside the \tablecaption.

\clearpage

\begin{deluxetable}{rrrcrrcr}
%\tablecolumns{8}
%\tablewidth{0pc}
\tablecaption{Photometry of artificial stars in the central part of ESO 410-G005
\label{tbl_1}}
\tablehead{ 
 \multicolumn{2}{c}{Input} & & &
\multicolumn{4}{c}{Recovered} \\
\cline{1-2} \cline{5-8} \\
\colhead{$I$} & \colhead{$V-I$}   & \colhead{}    & \colhead{Detection rate} &
\colhead{$\langle I\rangle$} & \colhead{$\sigma$}& \colhead{$\langle V-I\rangle$} & \colhead{$\sigma$}}
\startdata 
22.41 & 1.27& &  0.93&  22.34&0.17&  1.23&0.10 \\
23.12 & 1.10& &  0.90&  23.02&0.26&  1.06&0.13     \\
23.62 & 1.00& &  0.86&  23.50&0.33&  0.96&0.14       \\
24.12 & 1.00& &  0.80&  24.03&0.31&  0.98&0.17         \\
24.62 & 1.00& &  0.59&  24.33&0.61&  0.98&0.32  \\
25.12 & 1.00& &  0.13&  24.24&0.84&  1.02&0.32  \\

\enddata 
\end{deluxetable}

\begin{deluxetable}{lc}
\tablecaption{Properties of ESO\,410-G005
\label{tbl_2}}
\tablehead{
\colhead{Parameter}  &             } 
\startdata
  RA  (J2000.0)              &   $ 00^{\rm h}15^{\rm m}31.4$           \\
  Dec (J2000.0)              &    $-32\degr10'47''$               \\
  Galactic $l$   (deg)       &     357.85 \\
  Galactic $b$   (deg)        &    $-$80.71                   \\
  SuperGalactic $b$ (deg)     &   $-$0.26                     \\
  Dimension ($\arcmin$)         &            1.3$\times$1.0        \\
  $B_T$       (mag)       &              15.12            \\
  $V_T$       (mag)       &              14.35   \\
  $E(B-V)$    (mag)       &               0.013              \\
  Extinction: $A_B, A_I$ (mag)  &            0.06, 0.03             \\
  $V(<35\arcsec$) (mag)             &             14.80               \\
  $(V-I)_T$       (mag)   &             0.90$\pm$0.05                \\
  $\mu_{V,0}$   (mag arcsec$^{-2}$)  &   22.7$\pm$0.1               \\
  Scale length ($\arcsec$)        &      14.5$\pm$0.5                   \\    
  $I_{\rm TRGB}$  (mag)          &           22.4$\pm$0.15                 \\
  $(m-M)_0$  (mag)          &           $26.42\pm0.2$                 \\
  Distance (Mpc)            &           1.92$\pm$0.19             \\
  $(V-I)_{0,-3.5}$   (mag)    &           1.30$\pm$0.11         \\
  ${\rm [Fe/H]}$ (dex)            &           $-1.8\pm0.4$            \\
  Linear scale length (kpc)    &        0.13$\pm$0.01              \\
  $M_{B,0}$ (mag)             &            $-11.36\pm0.2$            \\
  Standard linear diameter  (kpc)  &       0.72                         \\
  Type                    &             dSph           \\
  Number of globular clusters    &         0             \\
  Projected separation from NGC\,55 (kpc) &   230      \\
  Radial distance to NGC\,55 (kpc)  & $340\pm250$ or $615\pm160$  \\            
\enddata
\end{deluxetable}

\begin{deluxetable}{ccccc}
\tablecaption{Global parameters of background galaxies on the eastern side
of ESO\,410-G005.
\label{tbl_3}}
\tablehead{
\colhead{Object} & \colhead{$V_T$} &  \colhead{$(V-I)_T$} & 
\colhead{$\mu_{V,0}$} &    \colhead{$R_{1/2}$}} 
\startdata
  1  &    20.0&     0.85 &       20.8&         $0\farcs82$ \\
  2  &    20.9&     1.20 &       21.8&         $0\farcs79$   \\
  3  &    21.1&     0.58 &       20.7&         $0\farcs41$     \\
  4  &    21.4&     1.20 &       21.8&         $0\farcs67$       \\ 
\enddata
\end{deluxetable}

\begin{deluxetable}{lcrrrccc}
\tablecaption{Distances and velocities of 13 galaxies in the Sculptor 
group\tablenotemark{1}.
\label{tbl_4}}
%\footnotesize
%\tiny
\tablehead{
\colhead{Name} & \colhead{RA} & \colhead{Dec} & \colhead{Type}
& \colhead{$B_T$} & \colhead{$V_0$} & \colhead{$D$} & \colhead{Note} \\
        &  \multicolumn{1}{c}{(J2000.0)}  & \multicolumn{1}{c}{(J2000.0)}  & 
& \multicolumn{1}{c}{[mag]}& \multicolumn{1}{c}{[km s$^{-1}$]} & 
\multicolumn{1}{c}{[Mpc]}    &      } 
\startdata
 UGCA 438          & 23 26 27.8 & $-$32 23 26&  10 &  14.07&    99&   2.08$\pm$0.12  & \tablenotemark{2} \\
 PGC 72228         & 23 43 45.9 & $-$31 57 33&   9 &  13.58&   299&               &                  \\
 NGC 7793          & 23 57 49.4 & $-$32 35 24&   8 &   9.70&   252&   3.27$\pm$0.08  &\tablenotemark{3,4} \\
 ESO 349-31 = PGC 621 & 00 08 13.2 & $-$34 34 42&  10 &  15.56&   216&   2.6:$\pm$0.8:  &\tablenotemark{5,6} \\
 NGC\,55            & 00 15 08.4 & $-$39 13 14&   9 &   8.39&   106&   1.66$\pm$0.20  &\tablenotemark{3} \\
 ESO 410-5 = kk003  & 00 15 31.4 & $-$32 10 48&  -3 &  14.85&      &   1.92$\pm$0.09  &\tablenotemark{7} \\
 Scl 22            & 00 23 51.7 & $-$24 42 18&  -3 &  17.73&      &   2.67$\pm$0.16  &\tablenotemark{8}  \\
 ESO 294-10 = PGC 1641 & 00 26 33.3 & $-$41 51 19&  10 &  15.60&    81&   1.71$\pm$0.07  &\tablenotemark{8}  \\
 NGC 247           & 00 47 08.5 & $-$20 45 36&   7 &   9.64&   215&   2.48$\pm$0.15  &\tablenotemark{3,4} \\
 NGC 253           & 00 47 34.2 & $-$25 17 32&   5 &   7.92&   281&   2.77$\pm$0.13  &\tablenotemark{3} \\
 ESO 540-30 = kk009 & 00 49 21.0 & $-$18 04 28&  -3 &  16.37&      &   3.19$\pm$0.13  &\tablenotemark{8} \\
 ESO 540-32 = kk010 & 00 50 24.5 & $-$19 54 25&  -3 &  16.44&      &   2.21$\pm$0.14  &\tablenotemark{3} \\
 NGC\,300           & 00 54 53.7 & $-$37 40 57&   7 &   8.79&   112&   2.10$\pm$0.10  &\tablenotemark{9,10} \\
\tablenotetext{1}
{[$22\fh5< {\rm RA} < 01\fh5$, $-15\degr < {\rm Dec} < -50\degr$] 
with $D < 4$ Mpc or $V_0 < 300$ km s$^{-1}$.}
\tablenotetext{2}{Lee \& Byun (1999)}
\tablenotetext{3}{Puche \& Carignan (1988)}
\tablenotetext{4}{Tammann (1987)}
\tablenotetext{5}{Laustsen et al.\ (1977)}
\tablenotetext{6}{Heisler et al.\ (1997)}
\tablenotetext{7}{Present work}
\tablenotetext{8}{Jerjen et al.\ (1998)}
\tablenotetext{9}{Freedman et al.\ (1992)}
\tablenotetext{10}{Soffner et al.\ (1996)}
\enddata
\normalsize
\end{deluxetable}

%% The following command ends your manuscript. LaTeX will ignore any text

\end{document}